# Gravitational 'Convergence' and Galaxy Cluster Masses

Tom Broadhurst
Department of Astronomy,
Campbell Hall, University of California, Berkeley USA

**ABSTRACT**

Two colour photometry of the cluster A1689 reveals a 'relative magnification-bias' between lensed blue and red background galaxies, arising from a dependence of the faint galaxy count-slope on colour. The colour distribution is skewed blueward of the far field, allowing us to measure the cluster magnification and to understand the notorious blueness of large arcs. We show that the magnification information can be combined with the usual image distortion measurements to isolate the local 'convergence' component of lensing and hence derive the projected mass. This is achieved through a simple local relation between the convergence and the observables, which can be applied generally over the surface a cluster. In the weak lensing limit, the convergence reduces to a dependence on the magnification alone, so that in the outskirts of clusters the surface-density of matter is obtained directly from the surface-density of background galaxies. Hence, useful lensing work requires colour information but not necessarily good seeing. Interestingly, convergence varies slowly at high redshift, saturating at a level depending on the Horizon distance, allowing a useful model-independent measurement of the Global Geometry.

*Subject headings:* clusters: Dark matter, — gravitational lensing

## 1. Introduction

The statistical distortion of lensed images pioneered by Tyson *et al* (1990) is firmly established as a tool for investigating the mass distribution of galaxy clusters (Bonnet *et al*, Fahlman *et al* 1994, Kaiser & Squires 1995, Tyson & Fischer 1995). Distortions are now understood to relate to the local gradient of the mass distribution, in the limit of weak lensing (Kaiser & Squires 1993). This means that in the outskirts of clusters the projected mass can be determined up to a constant of integration, yielding a firm lower limit. In the more strongly lensed central region of the cluster, the use of lensed distortions is more limited, as here the gradient of the mass distribution is uncertain by an unknown constant (Kaiser 1995, Schneider & Seitz 1995).

In this paper we point out the benefits of using magnification information for solving the above ambiguities, leading to absolute mass determinations. Firstly, we show how magnification may be measured in practice §1, using just the background galaxy counts behind A1689. In §2 we derive a general expression for the 'convergence' of a lens in terms of the observed magnification and distortion, to give the mass distribution. We compare the significances of distortion and magnification measurements and suggest an approriate observing strategy. Finally, in §3 we investigate the dependence of the convergence on the Global Geometry.

## 2. Detecting Gravitational Magnification

The effect of a lens is to bend passing light towards the center of mass, so the observer sees images pushed outward on the sky from the lens. Hence, compared to an unlensed patch of sky, the surface density of lensed galaxies is *reduced* by a factor equal to the local magnification for galaxies selected to a fixed isophote (surface brightness is conserved by lensing). In practice we cannot define a meaningful isophote, since, the majority of faint galaxies are too small to be well resolved, even with HST (see Kaiser Squires & Broadhurst 1995 - KSB). However, we may define a meaningful flux. In this case, the reduction in surface-density is countered by new galaxies brightened-up above the flux limit. The net effect on the lensed surface-density of background galaxies, $N'(<m)$, compared unlensed field, $N_o(<m)$, at fixed magnitude, $m$, has a sign and strength depending on the local magnification, $\mu$, and intrinsic count-slope, $S = dlogN(m)/dm$, simply:

$$\frac{N'}{N_o}(<m) = \mu^{2.5S-1} \qquad (1)$$

This is really a 'magnification-bias' (Turner *et al* 1984), for field galaxies. The bias is strongly *negative* for faint red galaxies, since the red count-slope $S_R \approx 0.15$, is much flatter than the lensing invariant slope $S = 0.4$, (Broadhurst 1994), i.e the sky expansion dominates over the increased visibility. For the bluest galaxies the magnification-bias is mildly positive $S_B \approx 0.5$, boosting the blue counts close to the lens.

The red counts provide a clean measure of the magnification, since it is always possible to define unambigously a background sample by selecting galaxies *redward* of the cluster E/S0 sequence. This follows simply from the fact that E/S0 galaxies are the reddest class of galaxies in the optical restframe, hence observed colours redder than the cluster E/S0's correspond to higher redshift, where larger K-corrections can make galaxies appear redder (for example, see the colour-redshift plane of faint field galaxies, Shade *et al*).

Furthermore, the cluster E/SO's are easy to identify by colour, forming a tight relation in the colour-magnitude plane (for example KSB Figure 3).

The lensing effect on the faint red counts is shown in Figure 1, for simple isothermal lenses. The surface-density of red galaxies is depressed at small radius, with a minimum at the critical radius, where the magnification is greatest. New data on A1689 (Broadhurst *et al* 1995) shows this signature clearly (plotted on Figure 1), demonstrating the feasibility of this approach. Notice, at small radius the magnification-bias changes sign, producing a net *excess* of red demagnified images. The converse is true for the blue counts, because they are steeper than the lensing invariant slope (eqn 1). Of course for blue galaxies, not all will be background, there will be some contribution from the foreground and cluster. However, this is probably unimportant for a low redshift cluster, depending on the poorly known faint end of the galaxy luminosity function. In principle then, the ratio of blue to red counts, $R = N_B(<m)/N_R(<m)$, provides a more sensitive measure of the magnification through a 'relative' magnification-bias:

$$\frac{R'}{R_o} = \mu^{2.5(S_B - S_R)} \qquad (2)$$

This index is of course independent of the coverage of background sky by the bright cluster galaxies, and has the further advantage of reduced sensitivity to the intrinsic angular clustering of field galaxies. This last point is irrelevant at faint magnitudes ($I > 24$) where the counts are essentially poisson on the scale of interest here (Efstathiou *et al* 1991, Giavalisco & Broadhurst 1995). For the count slopes quoted above, $R'/R_o \approx \mu$, a striking effect which is seen in the A1689 data, as shown in in Figure 2.

Also marked on Figure 2 is the "arc zone" where images are sufficiently magnified that they appear highly elongated. This is the region centered on the tangential critical radius, (Einstein ring) and as Figure 2 shows, it is the region where the ratio of blue/red galaxies is maximum due magnification bias. Therefore, in an absolute sense, we must expect arcs to be more commonly blue in colour than the field, to fixed magnitude, and is indeed a conspicous feature of cluster lensing (Tyson *et al* 1990). Radial arcs now being discovered in HST images and as Figure 2 shows these should also be relatively blue, being strongly magnified normal to the radial critical curve. However, in the central region the colours of de-magnified images ($\mu < 1$) will be *redder* than the field, as the magnification-bias switches sign (eqn 3), and this will be an interesting effect to look for.

When redshifts can be assigned to the faint background galaxy population a better measure of the magnification is achieved through $N(m, z)$, as all background galaxies can be made use of, not just those with extreme colours. Furthermore, the signal is better revealed in redshift since the competing effects of sky expansion and flux brightening scale with lens-source distance (Broadhurst, Taylor & Peacock 1995 - BTP). The magnification modified redshift distribution, to some fixed magnitude limit, $m_{lim}$, can be approximated as:

$$\frac{N'}{N_o}(z) = \mu(z)^{2.5\beta - 1} \qquad (3)$$

where the slope of the magnitude distribution, $\beta \approx dlogN(z| < m_{lim})/dm$, is an increasing function of redshift (BTP). This dependence arises because the magnitude limit moves up the galaxy luminosity function with increasing redshift. At high redshifts only the steep part of the luminosity function is detected, $\beta > 1$ producing a net excess. Conversely at lower redshift, $\beta < 1$ just behind the cluster, i.e. the sky expansion wins, depleting N(z). The overall effect is a 'stretching' of the lensed redshift distribution to higher redshift, compared with the field (BTP). At high spectral resolution the measurement of magnification would be improved by making use of $L - \Delta V$ relations, since $\Delta V$ is of course unaffected by lensing. This approach in now feasible as line widths (in emission and absorption) are now usefully resolved at magnitudes faint enough to be of interest.

Note, the obvious measure of magnification, namely the change in lensed image area at fixed isophote (first suggested by BTP), is unfortunately not a practical proposition, since as noted by KSB, the size distribution of faint galaxies seen with HST is intrinsically much broader than the lensing signal we seek, extending down to the diffraction limit.

## 3. Lens Convergence and Mass

Having demonstrated the way magnification may be obtained from observations of the background galaxy counts, we now address its role in determining the mass. We show that the magnification information complements the traditional image distortion measure in a very crucial way.

The effect of a lens can be separated into two independent parts (Young 1981), that due to the matter within a 'beam', (i.e. the integrated column of cluster matter covering a given patch of background galaxies) and that outside the beam, generated by the non-local anisotropy of the surrounding mass distribution. The local beam of matter purely magnifies images, preserving their shapes, and is referred to as the 'convergence', $\kappa$, of light i.e. a simple focusing of a light bundle. This is the quantity of interest for determining the mass distribution, since the local projected mass in this beam is simply equal to the convergence scaled by a geometric term, depending on the positions of lens and source along the 'optical bench'.

Unfortunately, we cannot just measure the magnification and relate it to the mass, since the anisotropy of the surrounding cluster mass generates a 'shear', $\gamma$, of the light bundle, further magnifying it. Therefore the shear and convergence must be decoupled to obtain the mass.

Remarkably, the convergence (or shear) can be isolated using two independent observables, image distortion: $\delta = a/b = (1 - \kappa + \gamma)/(1 - \kappa - \gamma)$ and magnification: $\mu = ab = 1/(1 - \kappa)^2 - \gamma^2$, where $a$ and $b$ are just stretch factors of the 'sky' along the the major and minor axes of the local shear direction. Solving for the convergence in terms of the observables gives:

$$\kappa = \Sigma \frac{4\pi G}{c^2} \frac{d_l d_{ls}}{d_s} = 1 \pm \frac{1 \pm |\delta|}{2\sqrt{\delta\mu}} \quad (4)$$

In general then, the convergence (and therefore mass) is only derivable with both distortion and magnification information. Indeed, it has been amply demonstrated that the convergence cannot be uniquely derived from the distortion field alone (Kaiser 1995, Schnieder & Seitz 1995). To convert the convergence to the local surface-density of mass, $\Sigma$, we need just the lens and source distances, $d_l, d_s$, where $d_{ls}$ is their separation.

We can only observe the modulus of the magnification and distortion, but the signs of the stretch factors change across the caustics leading to ambiguity unless the parity of an image can be recognised. The smallest solution ($\kappa < 1$, $\delta > 1$), is the most interesting, as this applies to the whole region outside the tangential critical curve (or Einstein ring). Between the radial and tangential critical curves one axis is reversed producing mirror images, so that $\delta$ and $\mu$ become defined negative. In this region there are two solutions separated by the $\kappa = 1$, contour or so-called 'critical surface density' (given by eqn 4) along which images are undistorted - being equally stretched in the tangential and radial directions (so that the magnification is not in general unity). Interior to the radial critical curve both axes are reversed, so that $\delta > 1$, requiring the largest of the four solutions, as shown in Figure 2. Note the bracketing of the $\kappa = 1$ contour between the critical curves is true in general (Kaiser 1995).

A general application of eqn 4 is certainly not practicable from the ground (as shown by radial span of the A1689 data in Figure 2), requiring we identify image parity and/or the two curves of infinite distortion and the distortionless $\kappa = 1$ contour which they bracket. Note that the simplest relation for convergence namely, $\kappa = 1 \pm \frac{1}{2|a|} \pm \frac{1}{2|b|}$, is unfortunately not a practical proposition for the reason we gave earlier regarding the difficulty of establishing image magnifications.

In the absence of any local mass ($\kappa = 0$) or void, the magnification and distortion are always related by $\mu = (1+\delta)^2/4\delta$, independent of distribution of the surrounding mass. Interestingly, for an axisymmetric void there is no net force on a photon, so $\mu = \delta = 1$ (Newtons' first theorem, in projection).

Importantly, in the weak lensing limit, the above expression for the convergence (eqn 4) reduces to a dependence on magnification *alone*:

$$\kappa \approx (\mu - 1)/2 \approx \frac{N'/N_o - 1}{5S - 2} \approx \frac{R'/R_o - 1}{5(S_B - S_R)} \tag{5}$$

In other words, in the outskirts of clusters, the local surface-density of galaxies can be simply converted into the local surface-density of mass. This is important to realise, since it implies useful lensing work can be carried out from the ground *without* good seeing!

## 4. Significance of Detection and Observing Strategy

The significance of the measurements of magnification and image distortion are closely related, as we now show.

The signal to noise of the distortion, $S/N_\delta$, in a circular bin of radius, $\Delta r$, whose center is separated from that of the cluster by, r, relates to the detection of the magnification, $S/N_\mu$, and to an isothermal mass through:

$$S/N_\delta \approx S/N_\mu \frac{\bar{\delta}}{\Delta\delta} \frac{\sqrt{n_o/n_c}}{|2.5S_c - 1|} \approx 6\frac{\bar{\delta}}{\Delta\delta} \frac{\Delta r}{r} \frac{d_{ls}}{d_s} \Big(\frac{\sigma}{1000km/s}\Big)^2 \sqrt{\frac{n_o}{50/\Box'}} \tag{6}$$

The relative precision of the detection of $\delta$ and $\mu$ depends most strongly on resolution, through the ratio of mean field galaxy distortion $\bar{\delta}$ over the unlensed dispersion $\Delta\delta$. Tests on HST data degraded to ground based seeing indicates a ratio of $\approx 2$ in good seeing of 0.5", for galaxies large enough to be resolved, which in practical terms means $I < 24$, or equivalently a surface-density of $\approx 50$ galaxies /$\Box$'. At fainter magnitudes the quality drops off since as HST imaging reveals, the sizes of galaxies becomes very small, barely resolvable by $I = 27$. At faint limits then, it is clear the magnification will be more easily detectable as $\bar{\delta}/\Delta\delta \to 0$. But at the practical limit from the ground ($I \approx 24$) the magnification is $\approx 1/4$ as precise as the distortion measurement through the combination of resolution and the restriction of the magnification measurement to the fraction of galaxies with a steep or flat count slope. For red galaxies, $|2.5S - 1| \approx 0.7$, with a surface-density, $n_c$, comprising $\sim 1/3$ of all galaxies.

Hence, for an isothermal mass distribution the detection for $I < 24$ selected galaxies is $\approx 12\Delta r/r$ in the measurement of distortion, allowing a $3\sigma$ detection in a bin of $\approx r/4$, or for the magnification the same level of detection is achieved in a larger bin of $\approx r$ i.e $\approx 1/4$ of the resolution. The detection of magnification is enhanced by going to fainter limits and, as discussed earlier, is also improved with redshifts. Currently the data on A1689 leads to an uncertainty in convergence measured over the whole field (8'×8') of only $\approx 5\%$, based on fairly modest depth photometry (6.5hrs integration on a 3.5m in 0.8" seeing, Broadhurst *et al* 1995).

Interestingly, the lensing signal for the mass profiles which are isothermal or steeper is strongest with maximum lens-source separation - i.e. for a clusters at *low* redshift, where $d_{ls}/d_s \to 1$, as emphasised by Kaiser (1995) (and not with the lens halfway between us and the sources as common wisdom has it!). The main advantage of working at low redshift is of course higher spatial resolution in mass, but at the expense of increased telescope time, set by the available field of view.

The requirement of obtaining magnification information for unambiguous determination of the mass requires a modification of the usual observing strategy. Imaging in two passbands is required, preferably deeper in the bluer band so that the red galaxies be detected. By apportioning the telescope time between two bands we would obtain the magnification information at no significant cost to the traditional distortion measurement, which could be derived from the sum of all images, since image shape and seeing are not strong functions of wavelength. With redshift information the measurement of $\kappa$ is improved by providing the source distance and through an increased sensitivity to the magnification (as outlined in §3). Again one would select the red galaxies to ensure that only background galaxies are observed.

## 5. Constraining the Global Geometry

As shown above, $\kappa$ is the product of mass and distance (eqn 4). This scaling with distance is also true for the shear of course, being an equivalent expression of the distortion and magnification, but note, is it *not* generally true of the magnification and distortion taken individually. Convergence measurements for two or more source planes are independent of the projected mass and hence it is interesting to investigate its geometric dependence. The ratio of convergence for two background planes measured in the same areal bin tends quickly to the Horizon limit:

$$\frac{\kappa_1}{\kappa_2} = \frac{d_{ls_1}d_{s_2}}{d_{s_1}d_{ls_2}} \to \frac{\kappa_{z1}}{\kappa_\infty} = \frac{g(z_l)[2 - \Omega_0 + \Omega_0 z_{s_1}] - g(z_{s_1})[2 - \Omega_0 + \Omega_0 z_l]}{g(z_l)[2 - \Omega_0 + \Omega_0 z_{s_1} + (\Omega_0 - 2)g(z_{s_1})]}$$

where $g(z) \equiv \sqrt{1 + \Omega_0 z}$ (BTP). The ratio is very strong function of the lens redshift, as shown in Figure 3 (for various choices of $\Omega$ and $\Lambda$). Here the lower of the two redshift planes is chosen at $z = 0.8$ - corresponding the likely mean redshift of intermediate magnitude background sources (see KSB for extrapolations of the CFRS survey to I=24). The more distant plane is allowed to vary in redshift, since we have currently little idea about the redshift distribution of fainter galaxies.

Figure 2 reveals two interesting regimes. At low lens redshift, $z_l < 0.2$, there is little dependence of the relative convergence with source redshift as it saturates at low redshift. This behaviour is very nice for measuring mass as there is need not worry about N(z). For intermediate lens redshift, $z \approx 0.5$, a steep dependence of convergence on redshift develops, so that the convergence at $z > 2$ is more than double that at $z = 0.8$, tending rapidly to its maximum at the Horizon distance, independent of the source redshift. For the interesting ranges of $\Omega$ and $\Lambda$, the limiting convergence ratio spans $\approx 20\%$ in size, several times larger than the likely uncertainty in the convergence measurement for a single massive cluster (discussed in §3). It is therefore conceivable that useful constraints on the Global Geometry may be obtained, assuming that the faintest galaxies lie at $z > 2$. Note, with the lens at higher redshift the test becomes less sensitive again, as it effectively lies among the sources.

In the weak limit, $\mu$ and $\delta$ may be used independently to measure the Geometry through the approximations, $\kappa \approx (\mu - 1)/2$, and $\gamma = (\delta - 1)/2)$, as both $\kappa$ and $\gamma$ scale with $d_{ls}/d_s$.

The discovery of statisically large samples of distant X-ray clusters (Rosati *et al* 1994, Schindler *et al* 1995) provides massive targets suitable for a statistical approach to this geometric measurement.

## 6. Conclusions

We have pointed out that blue and red faint field galaxies can undergo a significant relative magnification-bias when lensed, favouring the detection of blue over red galaxies in a magnitude limited sample. We have argued that magnification information is very important, since when combined with the usual image distortion measurements we may isolate the lens convergence and hence the mass of the lens. We have shown how to relate the convergence to local measurements of distortion and magnification through a simple expression which can be applied generally. To take advantage of the magnification information requires a different approach to observing. Imaging in two passbands is needed, preferrably deeper in the shorter wavelength band to detect the redder galaxies. Red galaxies are very useful for measuring the magnification bias, having a relatively shallow count slope and by virtue of being unambiguously in the background - for colours redder than the cluster E/SO sequence. In the weak limit, corresponding to the outskirts of the cluster, good seeing is not required to measure the mass, as the convergence relates only to the magnification and hence just to the counts. The first results of combining the distortion and magnification fields of the cluster A1689 using the method described here is very encouraging (Broadhurst *et al* 1995).

I am very grateful to Nick Kaiser for guidence and suggestions. I would also like to thank Steph Cote, Gordon Squires, Alex Szalay, Andy Taylor and Jens Villumsen for useful conversations and comments.

## REFERENCES

Bonnet, H., Mellier Y.,Fort B., 1994 ApJ, 427 L83

Broadhurst, T.J., 1994 proceedings of 5th Maryland meeting "Dark Matter" AIP 336 College Park, eds Holt. S., Bennett C.,L.

Broadhurst,T.J., Taylor A.N., Peacock J., 1995 ApJ, 438,49

Broadhurst, T.J., Kaiser N., Squires, G., Szalay A.S., Moller,P., 1995 ApJin prep.

Efstathiou, G.,Bernstein, G., Tyson, J., Katz, N., Guhathakurta, P.,1991 ApJ, 380, L47

Fahlman, G.,Kaiser, N.,Squires, G.,Woods, D. 1994 ApJ, 437,49

Giavalisco, M., Broadhurst T.J., 1995, ApJ, in prep.

Kaiser, N. 1995 ApJ1995 439, L1

Kaiser, N., Squires 1993 ApJ, 404,441

Kaiser, N., Squires G., Broadhurst, T.J., 1995 ApJ, 449, 460

Rosati, P., Dela Ceca, R., Burg, R., Norman, C., Giacconi, G., 1995, ApJ, 445, L11

Schindler *et al* 1995 A&A, in press

Schneider P., Seitz C., 1995 A&A, 297,287

Schade D., *et al*, 1995, Preprint

Turner E.D., Ostriker J.P., Gott J.R.III, 1984, ApJ, 284, 1

Tyson A., Fischer 1995 ApJ, in Press

Tyson A., Valdes, F., Wenk, R. A. 1990 ApJ, 349, L1

Young, P., 1981 ApJ, 244, 756

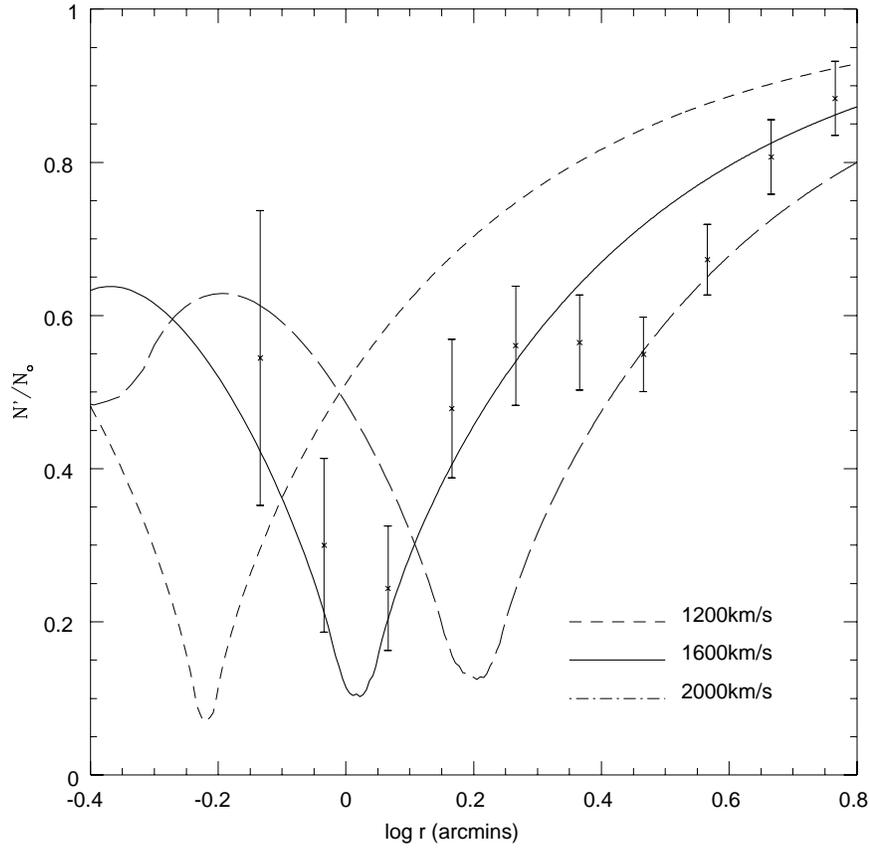

**Figure 1:** Radial behaviour of the magnification-bias, $N'/N_o$, for galaxies redward of the cluster E/OS sequence ($I < 24$, V-I $> 2.0$) of intrinsic count- slope $S = 0.15$, lensed by an isothermal mass with small core (core radius $<$ critical radius). Note the strong dependence on cluster mass ($\sigma^2$), and the pronounced minimum of the counts at the critical radius, where the magnification is greatest. The red counts behind A1689 are shown for comparison, clearly displaying the expected negative magnification-bias due to lensing.

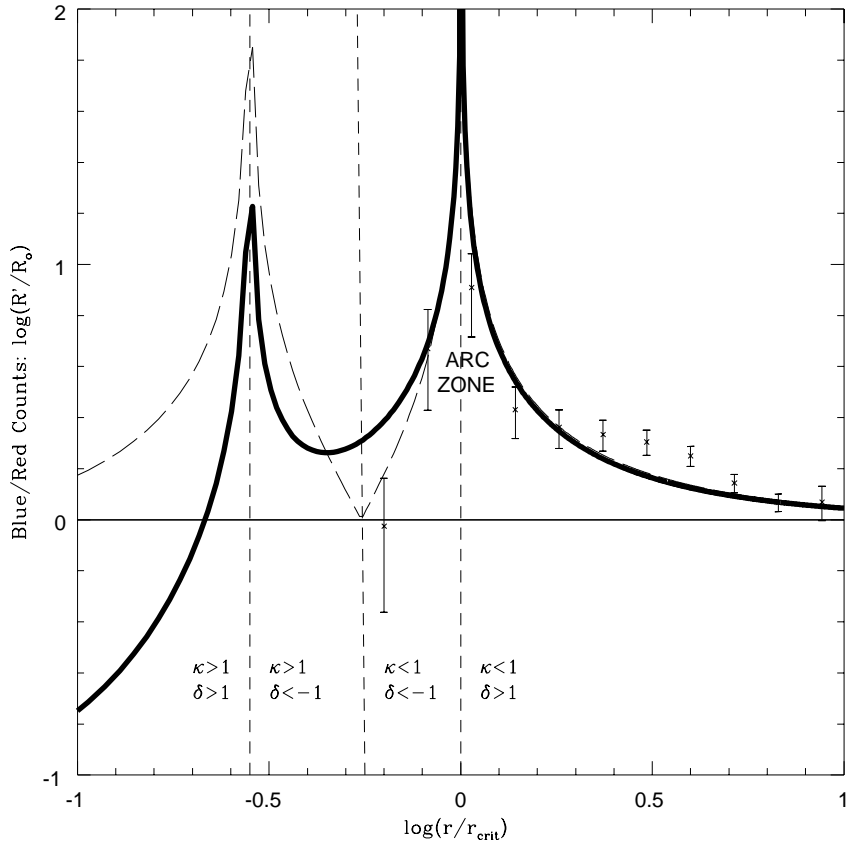

**Figure 2:** Radial behaviour of the relative magnification-bias (bold curve) of faint blue/red galaxies in units of the critical radius matched to data of A1689. The mass chosen here for demonstration is isothermal with a small core. The regions corresponding to the 4 solutions of the convergence relation (see text) are indicated, separated by vertical lines, corresponding to the inner (radial) critical curve, the distortionless $\kappa = 1$ radius and the outer (tangential) critical curve, or Einstein radius. The behaviour of the observable distortion, $|a/b|$, (dashed curve) shows the minimum at the radius where $\kappa = 1$. The maxima of blue/red lensed galaxies occur at the critical radii where the magnification is greatest (marked 'arc zone' and defined by the obvious arcs in the A1689 data, denoting the critical radius), naturally explaining why we should expect the giant tangential arcs to be blue relative to the unlensed far field.

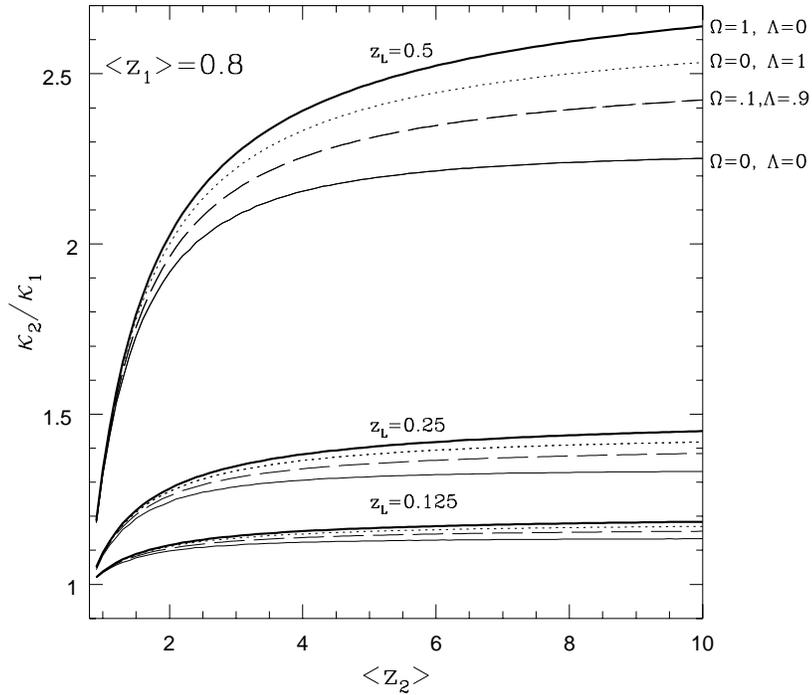

**Figure 3:** Behaviour of the relative convergence between sources at $<z_1> = 0.8$ and more distant sources at $<z_2>$, for three choices of lens redshift, $z_L$. In each case, the upper (bold) curve is for $\Omega = 1$ and the lower lighter curve for $\Omega = 0$. The figure also shows the full range of convergence ratio for flat univeres's, which spans $0.1 < \Omega < 1.0$ with $\Omega + \Lambda = 1$. There is some degeneracy, so that in the range $0.4 < \Omega < 1.0$, curves look very similar to combinations of $\Omega + \Lambda = 1$. Note the very strong dependence on lens redshift, so that at low redshift the source distribution and cosmology are unimportant for the measurement of mass. At higher lens redshift, the geometry may be explored without the knowledge of the uncertain redshift distribution, so long as the faint sources are distant $(z > 2)$.